\documentclass[aip, cha, amsmath, amssymb, twocolumn]{revtex4}
\usepackage{graphicx}
\usepackage{soul}
\usepackage{color}
\usepackage{amsmath}
\usepackage{pgf}
\usepackage{amsfonts}
\usepackage{graphicx}
\usepackage{soul}
\usepackage{float}
\usepackage{romannum}
\begin{document}
\title{Synchronization transitions in a hyperchaotic SQUID Trimer} 
\author{J. Shena$^{1}$, N. Lazarides$^{2}$, J. Hizanidis$^{1,2}$}
\affiliation{
 $^{1}$National University of Science and Technology "MISiS", Leninsky Prospect 
       4, Moscow, 119049, Russia \\
 $^{2}$Department of Physics, University of Crete, P. O. Box 2208, 71003 
       Heraklion, Greece
}
\date{\today}
\begin{abstract}
The phenomena of intermittent and complete synchronization between two out of 
three identical, magnetically coupled SQUIDs (Superconducting QUantum 
Interference Devices) are investigated numerically. SQUIDs are highly nonlinear 
superconducting oscillators/devices that exhibit strong resonant and tunable 
response to applied magnetic field(s). Single SQUIDs and SQUID arrays are 
technologically important solid state devices, and they also serve as a testbed 
for exploring numerous complex dynamical phenomena. In SQUID oligomers, the 
dynamic complexity increases considerably with the number of SQUIDs. The SQUID
trimer, considered here in a linear geometrical configuration using a realistic 
model with accesible control parameters, exhibits chaotic and hyperchaotic 
behavior in wide parameter regions. Complete chaos synchronization as well as 
intermittent chaos synchronization between two SQUIDs of the trimer is 
identified and characterized using the complete Lyapunov spectrum of the system 
and appropriate measures. The passage from complete to intermittent 
synchronization seems to be related to chaos-hyperchaos transitions as has been 
conjectured in the early days of chaos synchronization.
\end{abstract}
\pacs{05.45.+ b,05.45.Xt,84.30.-r}
\keywords{coupled nonlinear oscillators, SQUID trimer, chaos, hyperchaos, 
          chaos synchronization}
\maketitle
{\bf 
The phenomenon of synchronization between coupled and potentially chaotic 
oscillators is fundamental in nonlinear dynamics. Several types of 
synchronization of such oscillators, such as complete synchronization, phase 
synchronization, lag synchronization, rhythm synchronization, and generalized 
synchronization, have been described theoretically and observed experimentally. 
Chaos synchronization takes place in numerous physical and biological processes. 
The latter, in particular, seems to play an important role in the ability of 
biological oscillators, such as neurons, to act cooperatively. Chaos 
synchronization is a topic of current interest not only for its fundamental 
importance in nonlinear dynamics but also for its applicability in the context 
of electronic circuits, secure communications, and laser dynamics, 
among others. Generally speaking, chaos synchronization refers to a dynamical 
process in which coupled chaotic oscillators adjust a given measurable property 
of their dynamics to a common behavior, ranging from complete coincidence of 
trajectories to a functional relation between them. Another interesting type of
chaos synchronization is {\em intermittent synchronization}, in which temporal 
intervals of synchronization are interrupted by desynchronized activity. As an 
example, consider a system of three SQUIDs (Superconducting QUantum Interference 
Devices), that exhibits chaotic behavior for a wide range of parameters. In a 
chaotic state, two of the SQUIDs may be completely synchronized or the 
synchronization between them may be {\em intermittent}. Interestingly, 
{\em intermittent chaos synchronization} in the SQUID trimer is also related to 
the emergence of hyperchaos, as it has been conjectured in the past. These 
effects are explored here for the SQUID trimer using a numerical approach.     
}
\section{Introduction}
Chaos synchronization in coupled nonlinear systems 
\cite{Pecora1997,Fermat1999,Pikovsky2003,Anishchenko2014} 
has become a topic of great interest since 1990 \cite{Pecora1990}, 
based on earlier pioneering works \cite{Fujisaka1983, Afraimovich1986}. 
That interest stems not only from fundamental concerns of nonlinear dynamics, 
but mostly due to the possibilities that emerge for practical applications in 
electronic circuits 
\cite{Chua1992,Anishchenko1992,Chua1993,Rulkov1996,Rulkov1997,Bilotta2014} 
and secure communications
\cite{Kocarev1992,Kocarev1995,Hizanidis2010}
or in modeling biological systems and perceptive processes 
\cite{Mosekilde2002,Nowotny2008}, as well as in coupled lasers systems 
\cite{Winful1990,Terry1999,Shena2020b}.

A large number of experimental and theoretical works have addressed  chaos 
synchronization in Lorenz systems \cite{Lu2002,Sanchez2006}, coupled Duffing 
systems \cite{Wembe2009}, coupled chaotic oscillators by local feedback 
injections \cite{Yang1998}, two diffusively coupled Chua oscillators 
\cite{Dana2006}, coupled R\"ossler systems \cite{Hramov2004}, etc. Moreover,
synchronization of switching processes in coupled Lorenz systems 
\cite{Anishchenko1998b}, synchronization of chaotic oscillators by periodic 
parametric perturbations \cite{Anishchenko1997c}, loss of chaos synchronization 
through a sequence of bifurcations \cite{Anishchenko1997b} as well as the effect
of parameter mismatch on the mechanism of chaos synchronization loss 
\cite{Anishchenko1998}, and the influence of chaotic synchronization on mixing 
in the phase space of interacting systems \cite{Anishchenko2013}, have been 
investigated thoroughly by Vadim S. Anishchenko and his collaborators, who also 
proposed an indicator of chaos synchronization in Ref. \cite{Anishchenko2014b}. 
A particular type of chaos synchronization that has been also considered, 
although less often than other types, is that of {\em intermittent chaos 
synchronization} \cite{Heagy1995,Cenys1996,Chern1996,Gauthier1996,Baker1998,
Blackburn2000,Rim2001,Zhao2005,Kyprianidis2010,Cho2018,Shariff2018} where time 
intervals with exact coincidence of the trajectories of at least two oscillators 
of the considered system is interrupted by time intervals of asynchronous chaotic 
dynamics.

Here, a system of three identical SQUID oscillators arranged on a linear array,  
where the acronym stands for ``superconducting quantum interference device'', is 
investigated with respect to complete and intermittent chaos synchronization 
effects. In what follows, complete synchronization is meant to be the exact
coincidence of the trajectories of the SQUID oscillators at the two ends of the 
array. It is demonstrated numerically that in this sense, complete 
synchronization appears for wide parameter intervals. Furthermore, transitions 
from complete to intermittent chaos synchronization and vice versa are observed, 
which relate to corresponding chaos to hyperchaos transitions and vice versa 
\cite{Kapitaniak1995,Kapitaniak2005}. These effects are analyzed using the 
Lyapunov spectrum together with appropriate measures. The SQUID is a highly 
nonlinear superconducting oscillator/solid-state device, that responds 
resonantly to applied magnetic field(s). SQUIDs and SQUID oligomers, i.e., SQUID 
systems comprising a few SQUIDs, exhibit very rich dynamical behavior, including 
``snaking'' resonance curves, complex bifurcation structure, and chaos 
\cite{Hizanidis2018,Shena2020}. Specifically, the existence of homoclinic chaos 
in a pair of  SQUIDs has been shown theoretically 
\cite{Agaoglou2015,Agaoglou2017}. SQUIDs have been also used in large arrays to 
form metamaterials (SQUID metamaterials) that exhibit extraordinary properties 
investigated both theoretically and experimentally \cite{Lazarides2018} (and 
references therein).

\section{SQUID Trimer Model Equations} 
The simplest version of a SQUID consists of a superconducting ring that is 
interrupted by a Josephson junction (JJ) \cite{Josephson1962}. The latter is an
important nonlinear element in superconducting electronics, which, in its ideal
form is characterized by its critical current $I_c$ and a current-voltage curve 
which is given by the celebrated Josephson relations. A more realistic junction
model comprises three parallel branches; the one of them contains an ideal JJ, 
while the other two contain a resistor $R$ and a capacitor $C$. This is the so
called resistively and capacitively shunted junction (RCSJ) model for a 
realistic JJ, which has been used widely in theoretical and numerical studies.

\begin{figure}
   \includegraphics[scale=0.6]{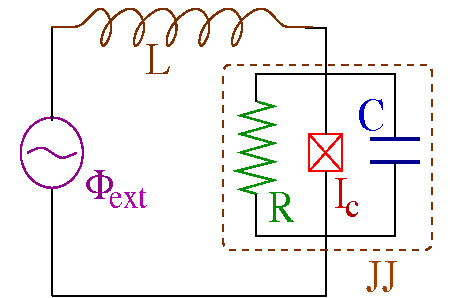}
\caption{Equivalent electrical circuit model for a single SQUID that relies on 
         the resistively and capacitively shunted junction (RCSJ) model of the
         Josephson junction (inside the dashed brown curve).}
\label{fig1}
\end{figure}

By employing the RCSJ model, connected in series with an inductance $L$ (due to
the SQUID ring) and a flux source $\Phi_{ext}$, an equivalent electrical circuit
model for the SQUID can be constructed, which is shown in Fig. \ref{fig1}. The 
external flux $\Phi_{ext}$, which often contains both constant (dc) and 
time-periodic (ac) components, is due to applied magnetic fields with 
appropriate orientation (usually perpendicular to the SQUID ring). The external 
flux induces currents in the SQUID ring due to Faraday's law, which in turn 
produce their own magnetic field along a direction opposite to that of the 
applied one. Thus, the flux which eventually threads the SQUID, $\Phi$, is the 
algebraic sum of the external flux $\Phi_{ext}$ and the flux due to the induced 
current, $L\, I_c$. This constitutes the flux-balance relation for a single 
SQUID. In any case, the dynamical equation for the flux $\Phi$ threading the 
SQUID loop can be obtained by direct application of Kirchhoff's laws to the 
equivalent electrical circuit for the SQUID in Fig. \ref{fig1}).

Consider three identical SQUIDS in an ``axial geometry'', i.e., a SQUID trimer, 
sush that their axes lie on the same line as in Fig. {\ref{fig2}. An externally 
applied magnetic field, whose direction is perpendicular to the rings of the 
SQUIDs (or equivalently is parallel to the SQUIDs' axes) so that magnetic flux 
$\Phi_{ext}$ threads their loop, induces currents in the SQUID rings through 
Faraday's law. These currents in turn produce their own magnetic field in each 
SQUID, whose magnetic flux threads the loops of the others. Thus, the SQUIDs are 
coupled together with magnetic dipole-dipole forces whose strength is quantified 
by their mutual inductance.

In order to derive the dynamical equations for the fluxes $\Phi_{n}$ ($n=1,2,3$)
threading the loops of the SQUIDs, we first write their flux balance relations 
\begin{eqnarray}
  \Phi_1 =\Phi_{ext} +L\, I_1 +M\, I_2 +\frac{M}{8} \, I_3, 
\nonumber \\
  \Phi_2 =\Phi_{ext} +M\, I_1 +L\, I_2 +M\, I_3, 
\label{eq01}
\\
  \Phi_3 =\Phi_{ext} +\frac{M}{8} \, I_1 +M\, I_2 +L\, I_3,
\nonumber
\end{eqnarray}
where $\Phi_n$ and $I_n$ is the flux threading the loop of the $n-$th SQUID and
the current flowing in the $n-$th SQUID, respectively, $L$ is the 
self-inductance of the SQUID ring (same for all three SQUIDs), and $M$ the 
mutual inductance between nearest-neighboring SQUIDs (i.e., between SQUIDs $1$ 
and $2$ and SQUIDs $2$ and $3$). Assuming that the strength of the dipole-dipole 
interaction between SQUIDs falls off as the inverse cube of their distance, we 
have adopted the value of $M/8$ for the coupling strength between SQUIDs $1$ and 
$3$.  
\begin{figure}
   \includegraphics[scale=0.4]{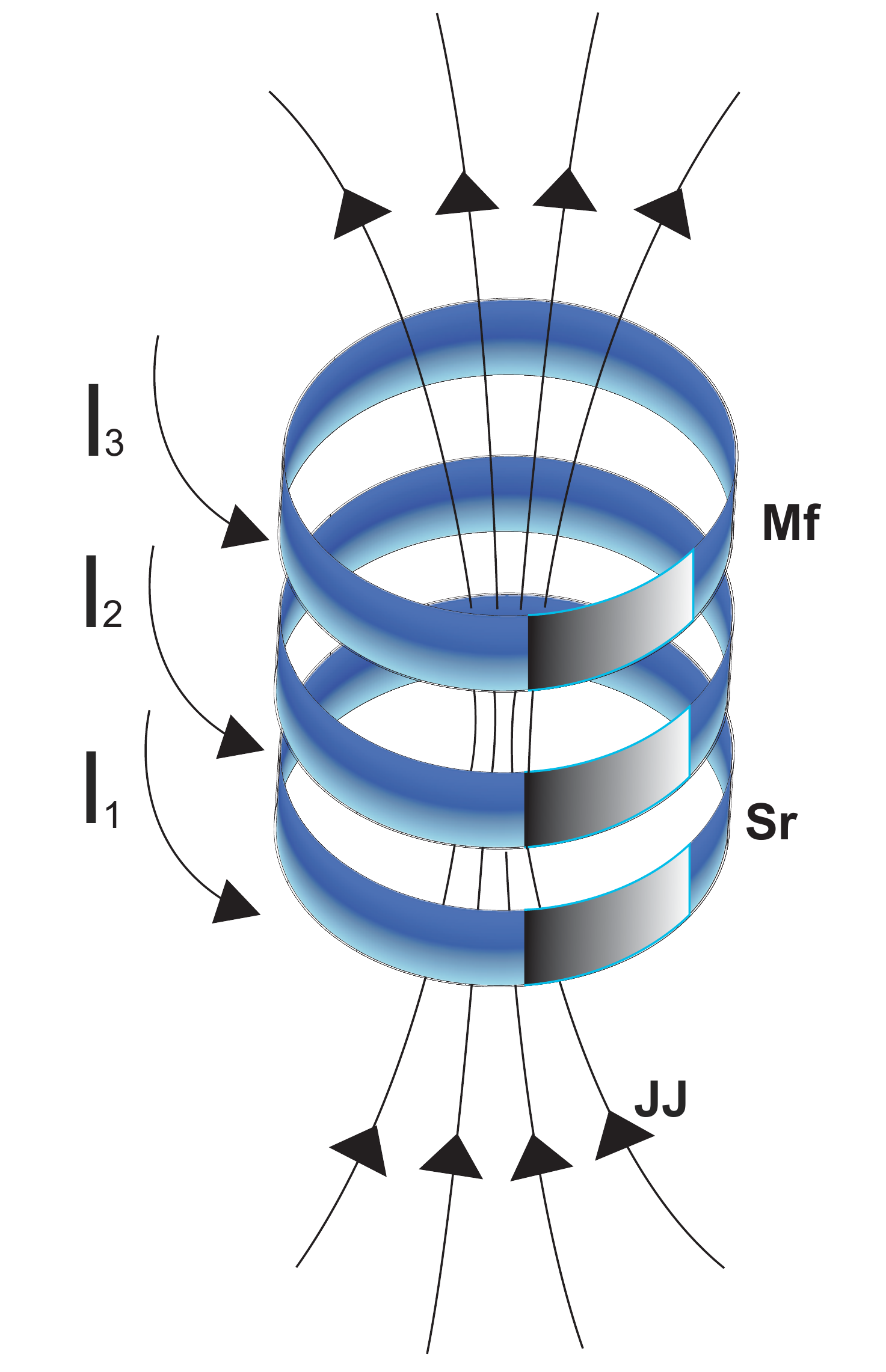}
\caption{Schematic diagram of a SQUID trimer in the ``axial geometry'' subject 
         to a magnetic field Mf, which are coupled together through their mutual 
         inductance with strength $\lambda$. 
         Sr is the superconducting ring, JJ is the Josephson Junction, 
         and $I_1$, $I_2$, and $I_3$ are the induced currents.} 
\label{fig2}
\end{figure} 

By dividing Eqs. (\ref{eq01}) with the self-inductance $L$, by rearranging terms,
and by defining the dimensionless coupling strength as
\begin{equation}
\label{eq02}
   \lambda =\frac{M}{L},
\end{equation}
Eqs. (\ref{eq01}) can be written in matrix form as
\begin{equation}
\label{eq03}
   {\bf \hat{\Lambda}} \vec{I} =\vec{\cal F},
\end{equation}
where
\begin{equation}
\label{eq04}
   {\bf \hat{\Lambda}}=
\begin{bmatrix}
   1 & \lambda & \frac{\lambda}{8} 
\\[2ex]
   \lambda & 1 & \lambda
\\[2ex]
   \frac{\lambda}{8} & \lambda & 1
\end{bmatrix}
, ~~\vec{I}=
\begin{bmatrix} 
I_1 \\[2ex]
I_2 \\[2ex] 
I_3 
\end{bmatrix}
, ~~\vec{\cal F}=\frac{1}{L}
\begin{bmatrix} 
   \Phi_1 -\Phi_{ext}
\\[2ex]  
   \Phi_2 -\Phi_{ext}
\\[2ex] 
   \Phi_3 -\Phi_{ext}
\end{bmatrix}
.
\end{equation}
The current flowing in the $n-$th SQUID is provided in terms of the flux 
$\Phi_n$ ($n=1,2,3$) threading its loop by the resistively and capacitively 
junction (RCSJ) model, as \cite{Likharev1986}
\begin{eqnarray}
\label{eq05}
  I_n =-C\, \frac{d^2\Phi_n}{dt^2} -\frac{1}{R}\, \frac{d\Phi_n}{dt} 
       -I_c\, \sin\left(2\pi\frac{\Phi_n}{\Phi_0}\right), 
\end{eqnarray}
where $\Phi_0$ is the flux quantum and $t$ is the temporal variable. 
By multiplying Eq. (\ref{eq03}) with the inverse of the matrix 
${\bf \hat{\Lambda}}$, and by substituting the components of $\vec{I}$ using
Eq. (\ref{eq05}), we get
\begin{widetext}
\begin{gather}
\label{eq06}
L
\begin{bmatrix} 
   C \frac{d^2 \Phi_1}{dt^2} +\frac{1}{R} \frac{d \Phi_1}{dt} 
   +\frac{2\pi}{\Phi_0} I_c\, \sin\left( 2\pi \frac{\Phi_1}{\Phi_0} \right) 
\\[2ex] 
   C \frac{d^2 \Phi_2}{dt^2} +\frac{1}{R} \frac{d \Phi_2}{dt} 
   +\frac{2\pi}{\Phi_0} I_c\, \sin\left( 2\pi \frac{\Phi_2}{\Phi_0} \right) 
\\[2ex] 
   C \frac{d^2 \Phi_3}{dt^2} +\frac{1}{R} \frac{d \Phi_3}{dt} 
   +\frac{2\pi}{\Phi_0} I_c\, \sin\left( 2\pi \frac{\Phi_3}{\Phi_0} \right) 
\end{bmatrix}
   =\frac{1}{D}
\begin{bmatrix}
   1-\lambda^{2} & -\lambda+\frac{\lambda^{2}}{8} & 
   \lambda^{2} -\frac{\lambda}{8}  
\\[2ex]
   -\lambda+\frac{\lambda^{2}}{8} & 1-\frac{\lambda^{2}}{64} &  
   -\lambda+\frac{\lambda^{2}}{8} 
\\[2ex]
   \lambda^{2} -\frac{\lambda}{8}  & -\lambda+\frac{\lambda^{2}}{8} & 
   1-\lambda^{2}
\end{bmatrix}
\begin{bmatrix} 
   \Phi_1 - \Phi_{ext} 
\\[2ex]  
   \Phi_2 - \Phi_{ext} 
\\[2ex] 
   \Phi_3 - \Phi_{ext} 
\end{bmatrix}
\end{gather},
\end{widetext}
where
\begin{equation} 
\label{eq07}
   D \equiv \det\left( {\bf \hat{\Lambda}}\right) 
    =1 -\frac{129}{64} \lambda^2 +\frac{1}{4}\lambda^3.
\end{equation}
In the following, the external flux is considered to be of the form 
\begin{eqnarray}
\label{eq08}
  \Phi_{ext} =\Phi_{dc} +\Phi_{ac} \, \cos( \omega t ),
\end{eqnarray}
i.e., it contains both a constant (dc) flux bias $\Phi_{dc}$ and an alternating 
(ac) flux of amplitude $\Phi_{ac}$ and frequency $\omega$.

Equations (\ref{eq07}) and (\ref{eq08}) are normalized using the relations
\begin{eqnarray}
\label{eq09}
  \phi_n =\frac{\Phi_n}{\Phi_0}, ~~~\phi_{ac,dc}=\frac{\Phi_{ac,dc}}{\Phi_0},
  ~~~\tau=\frac{t}{\omega_{LC}^{-1}}, ~~~\Omega=\frac{\omega}{\omega_{LC}},
\end{eqnarray}
where $\omega_{LC} =1 / \sqrt{L C}$ is the inductive-capacitive ($L\, C$) SQUID 
frequency. Eventually, the normalized equations read
\begin{widetext}
\begin{gather}
\label{eq10}
\begin{bmatrix} 
   \ddot{\phi}_1 +\gamma \dot{\phi}_1 +\beta \sin( 2\pi \phi_1 ) 
\\[2ex] 
   \ddot{\phi}_2 +\gamma \dot{\phi}_2 +\beta \sin( 2\pi \phi_2 ) 
\\[2ex] 
   \ddot{\phi}_3 +\gamma \dot{\phi}_3 +\beta \sin( 2\pi \phi_3 ) 
\end{bmatrix}
   =\frac{1}{D}
\begin{bmatrix}
   1-\lambda^{2} & -\lambda+\frac{\lambda^{2}}{8} & 
   \lambda^{2} -\frac{\lambda}{8}  
\\[2ex]
   -\lambda+\frac{\lambda^{2}}{8} & 1-\frac{\lambda^{2}}{64} &  
   -\lambda+\frac{\lambda^{2}}{8} 
\\[2ex]
   \lambda^{2} -\frac{\lambda}{8}  & -\lambda+\frac{\lambda^{2}}{8} & 
   1-\lambda^{2}
\end{bmatrix}
\begin{bmatrix} 
   \phi_1 -\phi_{ext}
\\[2ex]  
   \phi_2 -\phi_{ext}
\\[2ex] 
   \phi_1 -\phi_{ext}
\end{bmatrix}
\end{gather},
\end{widetext}
and
\begin{equation}
\label{eq11}
   \phi_{ext} (\tau) =\phi_{dc} +\phi_{ac} \cos( \Omega \tau ),
\end{equation}
where
\begin{equation}
\label{eq13}
   \beta=\frac{I_c L}{\Phi_0}, \qquad \gamma=\frac{1}{R} \sqrt{ \frac{L}{C} }
\end{equation}
is the rescaled SQUID parameter and the loss coefficient, respectively.
\begin{figure}[!t]
   \includegraphics[scale=0.44]{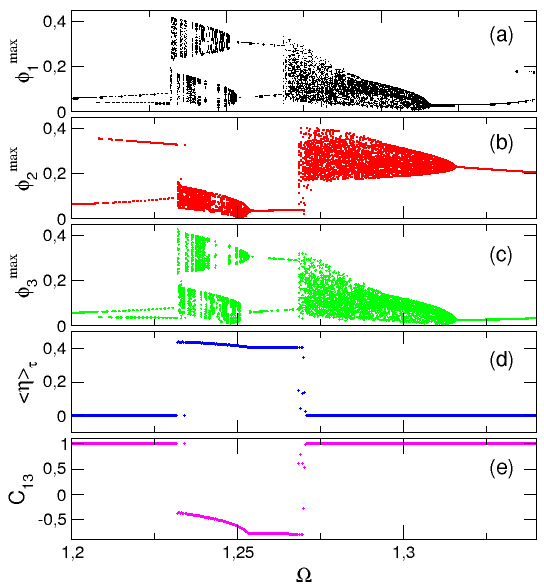}
\caption{Bifurcation diagrams of (a) $\phi_1$ (black), (b) $\phi_2$ (red), and 
         (c) $\phi_3$ (green), as a function of the driving frequency $\Omega$. 
         (d) and (e) are the averaged in time Euclidean distance 
         ${\langle \eta \rangle}_\tau$ and the correlation function $C_{13}$, 
         respectively. 
         Parameters: $\lambda=0.08$, $\phi_{ac}=0.02$, $\gamma=0.024$, 
         $\phi_{dc}=0$, and $\beta=0.1369$.}
\label{fig3}
\end{figure}
\begin{figure*}[!t]
   \includegraphics[scale=1]{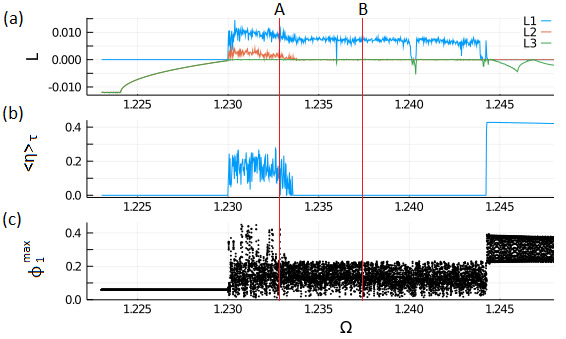}
\caption{(a) The three largest Lyapunov exponents (the rest are negative),
         (b) the average of $\eta$ over time, 
             $\left\langle \eta\right\rangle _{\tau}$, and
         (c) the maximum value of the magnetic flux $\phi_1$ in each driving 
             cycle, $\phi_1^{max}$, as a function of the driving frequency 
             $\Omega$. 
             Red line A corresponds to $\Omega = 1.233$ and 
             B to $\Omega = 1.2375$. We observe two main regions: The first one 
             is between $\Omega=1.23$ and $\Omega=1.234$ and corresponds to 
             intermittent hyperchaos synchronization  
 ($0.38 > \langle \eta \rangle _\tau > 0.01$, $L_1 > L_2 > 0, L_3 =0$) 
             while the second one lies between $\Omega=1.234$ and $\Omega=1.244$
             and corresponds to complete chaos synchronization  
   ($\left\langle \eta\right\rangle _{t} < 0.01$, $L_{1} > 0, L_2 =L_3 =0$). 
   Parameters: $\lambda=0.1075$, $\phi_{ac}=0.02$, $\gamma=0.024$, $\phi_{dc}=0$, 
   and $\beta=0.1369$.} 
\label{fig4}
\end{figure*}

\begin{figure*}[!t]
   \includegraphics[scale=1]{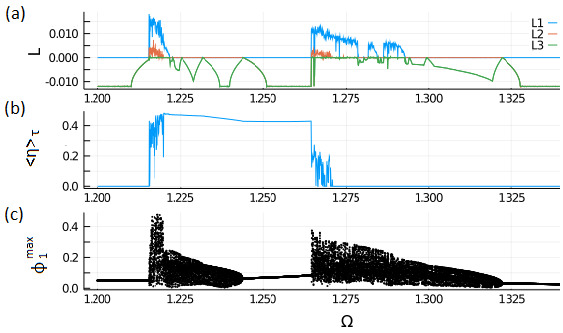}
\caption{(a) The three largest Lyapunov exponents (the rest are negative),
         (b) the average of $\eta(\tau)$ over time, 
             $\langle \eta \rangle _{\tau}$, and 
         (c) the maximum value of the magnetic flux in SQUID $1$ in each driving
             cycle, $\phi_1^{max}$, as
             a function of the driving frequency $\Omega$. 
             We observe three regions in which at least one Lyapunov exponent is
             positive: 
             The first one lies between $\Omega=1.271$ and $\Omega=1.293$, in 
             which the chaotic trajectories of SQUIDs $1$ and $3$ are completelly 
             synchronized 
 ($\langle \eta \rangle _{\tau} < 0.01$ and $L_1 > 0$, $L_2 =0$, $L_3=0$). 
 Note however in this region the existence of at least two narrow windows in 
 which the SQUID trimer exhibits periodic behavior along with synchronization of
 the trajectories between SQUIDs $1$ and $3$
 ($\langle \eta \rangle _\tau < 0.01$ and $L_1 =0$, $L_2, L_3 < 0$). 
             The second and the third ones lie in the frequency range 
             $(1.266, 1.271)$ and $(1.216, 1.22)$,
             respectively, in which the system exhibits hyperchaotic behavior
             ($L_{1} > L_{2} >0$, $L_3=0$). In the second region, intermittent 
             chaos synchronization is observed between SQUIDs $1$ and $3$
             ($0.01 < \langle \eta \rangle _{\tau} < 0.38$). In the 
             third region however, the situation is more complicated since the
             value of $\langle \eta \rangle _\tau$ often exceeds the limiting 
             value $0.38$ (see text).    
             Parameters: $\lambda=0.1075$, $\phi_{ac}=0.025$, $\gamma=0.024$, 
             $\phi_{dc}=0$, and $\beta=0.1369$.} 
\label{fig5}
\end{figure*} 

In what follows, the external dc flux is set to zero, i.e., $\phi_{dc} = 0$, for
simplicity. The values of the SQUID model $\beta$ and $\gamma$ are obtained from
Eqs. (\ref{eq13}), using the experimentally determined parameters for the 
equivalent circuit elements $L = 120 ~pH$, $C = 1.1 ~pF$, $R = 500 ~\Omega$, and 
$I_{c} =2.35 ~mA$ \cite{Zhang2015}. Using these values in Eqs. (\ref{eq13}) we get
$\beta =0.1369$ ($\beta_L \simeq0.86$) and $\gamma =0.024$, while the $L\, C$ 
frequency is $f_{LC} = \omega_{LC} / 2 \pi =13.9 ~GHz$.

Note that most of the numerical work below has been performed with Julia
programming language and the DynamicalSystems package \cite{Datseris2018}. The  
relevant codes used in this paper can be found in 
\url{https://github.com/Joniald/Squid_Trimer}.

\section{Chaos Synchronization and Quantitative Measures}
As mentioned earlier, complete or intermittent chaos synchronization between the
two SQUIDs at the ends of the trimer is observed, which we denote by $1$ and $3$. 
In order to quantify the synchronization between them, we adopt two different 
measures, namely, the instantaneous Euclidean distance:
\begin{equation} 
\label{eq14}
   \eta(\tau) =\sqrt{ \left[ \phi_1(\tau) -\phi_3(\tau) \right]^2 
                     +\left[\dot{\phi_1}(\tau)- \dot{\phi_3}(\tau) \right]^2 }, 
\end{equation} 
in the reduced phase space of SQUIDs $1$ and $3$, which is an intuitive measure 
of the quality of synchronization \cite{Baker1998}, averaged over a time-interval
$\Delta \tau$, $\langle \eta \rangle_\tau$, and the correlation coefficient 
between the normalized fluxes of SQUIDs $1$ and $3$:
\begin{equation} 
\label{eq15}
   C_{13} =\frac{ \langle [ \phi_1(\tau) -\mu_1 ][ \phi_3(\tau) -\mu_3 ] \rangle }
                { \sigma_1 \sigma_3 },  
\end{equation} 
where 
\begin{equation} 
\label{eq16}
   \mu_i =\frac{1}{\Delta \tau} \int_{\tau_{tr}}^{\tau_{tr} +\Delta \tau}
       \phi_i(\tau) d\tau \rightarrow \frac{1}{M} \sum_{n=1}^M \phi_i(\tau_n),
\end{equation} 
is the temporal average of $\phi_i(\tau)$ ($i=1,3$) over the time interval 
$\Delta \tau$, $\tau_{tr}$ is the time allowed for transients to die out, $M$ is 
the number of integration time-steps in $\Delta \tau$, and $\sigma_i$ are the 
standard deviations of $\phi_i$, given by:  
\begin{equation} 
\label{eq17}
   \sigma_i =\sqrt{ \langle [ \phi_i(\tau) -\mu_i ]^2 \rangle }.
\end{equation} 
Complete synchronization between the trajectories of SQUID $1$ and $3$ is 
achieved  when $\langle \eta \rangle_\tau =0$, and $C_{13} =1$. For 
intermittent chaos synchronization, $1 > \langle \eta \rangle_\tau > 0$ and 
$C_{13} < 1$. (The value of $C_{13} =-1$ if it ever occurs would indicate chaos 
anti-synchronization \cite{Kim2003}.) 

The SQUID trimer exhibits chaotic dynamics in relatively wide parameter 
intervals. Here we choose a value for the amplitude of the alternating (ac) flux 
$\phi_{ac} =0.02$ which is relatively low and certainly within the 
experimentally accesible values of this quantity. In Figs. \ref{fig3} (a)-(c), 
the bifurcation diagrams for the flux of all three SQUIDs are shown as functions 
of $\Omega$. Together we have plotted the averaged Euclidean distance 
$\langle \eta \rangle_\tau$ (Fig.~\ref{fig3} (d)) and the correlation function 
$C_{13}$ (Fig.~\ref{fig3} (e)). As it can be observed, the frequency interval in 
these figures is above the geometrical frequency $\Omega_{LC} =1$, and below the 
linearized SQUID frequency $\Omega_{SQ} =\sqrt{ 1 +\beta_L }$ ($\simeq 1.36$ for
the value of $\beta_L =0.86$ used in the simulations). The value of 
$\lambda=0.08$ used for obtaining the results are within the calculated ones in 
~\cite{Zhang2015} for this particular geometric configuration (e.g., the axial 
configuration). As it can be observed in all subfigures, there are several 
frequency intervals where the dynamics is chaotic in which SQUID $1$ and $3$ are 
completely synchronized ($\langle \eta \rangle_\tau =0$ and $C_{13} = 1$. There 
are also intervals in which intermittent chaos synchronization appears. In the 
latter, $1 > \langle \eta \rangle_\tau > 0$ and $C_{13} < 1$. Note that the 
two measures agree in all frequency intervals on whether SQUIDs $1$ and $3$ are 
completely or intermittently synchronized.

In order to identify precisely the frequency intervals in which the SQUID trimer 
exhibits chaotic dynamics, we calculate the full Lyapunov spectrum. A typical 
example is shown in Fig. \ref{fig4} for $\phi_{ac} = 0.02$ and 
$\lambda = 0.1075$, where the three largest Lyapunov exponents 
$L_1 > L_2 > L_3$ (Fig. \ref{fig4} (a)) are plotted together with 
$\langle \eta \rangle_\tau$ (Fig. \ref{fig4} (b)), and the amplitude of the 
flux threading SQUID $1$, $\phi_1^{max}$ (Fig. \ref{fig4} (c)). These results 
are obtained by initializing the SQUID trimer at $\Omega =1.233$ and performing 
a sweep in the decreasing direction in small steps 
$\Delta \Omega =2.5 \times 10^{-5}$. For each value of $\Omega$, except for the 
first one, the solution obtained for the previous value of $\Omega$ is set as 
initial condition for the SQUID trimer.

The three largest Lyapunov exponents are sufficient for characterizing the 
dynamics of the SQUID trimer, while the remaining $L_4$, $L_5$, $L_6$, and $L_7$ 
are always negative. Note that since time $\tau$ is treated as a dependent 
variable (and this is why there are seven Lyapunov exponents instead of six), 
one of the exponents is always zero. As the driving frequency $\Omega$ decreases 
from the maximum shown value $\Omega = 1.250$ down to $\Omega = 1.244$, it is 
observed that the two largest Lyapunov exponents $L_1$ and $L_2$ are zero while 
the third largest is mostly negative ($L_3 < 0$), indicating quasiperiodic 
dynamics. There is no synchronization between SQUID $1$ and $3$ in this dynamical 
state as can be inferred by the corresponding values of 
$\langle \eta \rangle_\tau \simeq 0.42$. Note that at a particular frequency 
$\Omega = 1.248$, the three largest Lyapunov exponents are all zero, i.e., 
$L_1 =L_2 =L_3 =0$, indicating a bifurcation from one quasiperiodic dynamical 
state to another one.

At approximately $\Omega = 1.244$, the SQUID trimer undergoes a quasiperiodicity 
to chaos transition and consequently the largest Lyapunov exponent $L_1$ becomes 
positive ($L_1 > 0$, blue curve) while $L_3$ becomes zero like $L2$ (green and 
orange curves). With further decreasing $\Omega$, the (positive) value of the 
largest exponent $L_1$ remains almost constant on average (note however the 
existence of a couple of narrow periodic windows where $L_1$ drops to zero)
until $\Omega$ reaches $1.234$. Note that in this frequency interval, i.e.,  
for $\Omega$ in $[1.244, 1.234]$, in which the dynamics of the SQUID trimer is 
chaotic, SQUIDs $1$ and $3$ are completely synchronized as can be inferred from 
the corresponding values of $\langle \eta \rangle_\tau \simeq 0$ there.

At $\Omega =1.234$, the second largest Lyapunov exponent $L_2$ becomes also 
positive, while $L_3$ remains zero, indicating a chaos to hyperchaos 
transition. That value of $\Omega$ also signifies a transition from complete to 
intermittent chaos synchronization between SQUIDs $1$ and $3$. The hyperchaotic 
dynamical state persists down to $\Omega =1.230$. In the frequency interval 
$[1.230,1.234]$, as mentioned earlier, SQUID $1$ and $3$ exhibit intermittent 
chaos synchronization. This can be inferred from the value of 
$\langle \eta \rangle_\tau$ which is clearly above zero and strongly fluctuating 
but less than the limiting value of $0.38$. Indeed, we have empirically found 
that for $\langle \eta \rangle_\tau \leqslant 0.38$ we have intermittent chaos
synchronization while for $\langle \eta \rangle_t > 0.38$ the two SQUIDs, $1$ 
and $3$, are neither synchronized together nor with SQUID $2$ (middle SQUID). 
Finally, for even lower values of the driving frequency $\Omega$, i.e., for 
values in the interval $[1.223, 1.230]$, a transition from a hyperchaotic to a 
periodic state occurs, and the maximum Lyapunov exponent becomes zero 
($L_1 =0, L_2, L_3 < 0$).

Thus, in the results shown in Fig. \ref{fig4}, we can identify four different 
types of behavior that mostly dominate the dynamics:

(i) $\Omega$ in $(1.244, 1.250]$: Quasiperiodicity ($L_1 =L_2 =0$, $L_3 < 0$)
without any type of synchronization between SQUID $1$ and $3$ 
($\langle \eta \rangle_\tau \simeq 0.42$).

(ii) $\Omega$ in $(1.234, 1.244]$:  Chaos ($L_1 > 0$, $L_2 =L_3 =0$), and 
complete chaos synchronization between SQUID $1$ and $3$ 
($\langle \eta \rangle_\tau < 0.01$). In practice, the values of 
$\langle \eta \rangle_\tau$ obtained in this dynamical state are all less than 
$0.01$. 

Chaotic behavior in a system whose Lyapunov spectrum has one positive and two
vanishing exponents, while all the others are negative, is referred to as 
``toroidal chaos'' in a recent classification of chaotic regimes
\cite{Letellier2021}.  

(iii) $\Omega$ in $(1.230, 1.234]$: Hyperchaos ($L_1 > L_2 >0$, $L_3 =0$), and 
intermittent chaos synchronization between SQUID $1$ and $3$
($0.38 > \langle \eta \rangle_\tau > 0.01$).

(iv) $\Omega$ in $[1.230, 1.223]$: Periodic dynamics 
($L_1 =0$, $L_2 < 0$, $L_3 < 0$), with SQUIDs $1$ and $3$ being synchronized 
($\langle \eta \rangle_\tau \simeq 0$).

By changing the parameters of the system we may observe a plethora of 
transitions between dynamical regions, as $\Omega$ is varied. To illustrate 
this, we have produced a similar plot to Fig.~\ref{fig4}, for $\lambda=0.1075$ 
and $\phi_{ac}=0.025$. The results are shown in Fig.~\ref{fig5}. Compared to 
Fig.~\ref{fig4}, the scenario here presents additional dynamical regions, and a 
total of six different types of behavior which are the following:

(i) $\Omega$ in $(1.322, 1.34]$: 
Periodic dynamics ($L_1 =0$, $L_2 < 0$, $L_3 < 0$), with synchronization between 
SQUIDs $1$ and $3$ ($\langle \eta \rangle_\tau <0.01$).

(ii) $\Omega$ in $(1.293, 1.322]$: Quasiperiodic dynamics 
($L_1 =L_2 =0, L_3 < 0$), with synchronization between SQUIDs $1$ and $3$ 
($\left\langle \eta \right\rangle_\tau < 0.01$).
In this interval, two bifurcations from a quasiperiodic state to another are
visible for the values of $\Omega$ at which $L_1 =L_2 =L_3 =0$.

(iii) $\Omega$ in $[1.271, 1.293]$: Chaos ($L_1 > 0$, $L_2 =L_3 =0$), and 
complete chaos synchronization between SQUIDs $1$ and $3$ 
($\left\langle \eta \right\rangle_\tau <0.01$). This is yet another case of
toroidal chaos \cite{Letellier2021} already mentioned in the discussion of Fig.
\ref{fig4}.

In this region of $\Omega$, there are also visible at least two windows in which 
the SQUID trimer exhibits periodic behavior with synchronized trajectories of 
SQUIDS $1$ and $3$. This dynamical behavior has been also observed in region (i) 
above, and it will not be further analyzed.   

(iv) $\Omega$ in $(1.266, 1.271]$:
Hyperchaos ($L_1 > L_2 >0$, $L_3 =0$), and intermittent chaos synchronization 
between SQUID $1$ and $3$
($0.38 > \left\langle \eta \right\rangle_\tau > 0.01$).

(iv) $\Omega$ in $(1.244, 1.266]$: 
Periodic dynamics ($L_1 =0$, $L_2 < 0$, $L_3 < 0$), without synchronization 
between SQUIDs $1$ and $3$ ($\left\langle \eta \right\rangle_\tau > 0.38$).

(v) $\Omega$ in $(1.22, 1.244]$: 
Quasiperiodicity ($L_1 =L_2 =0$, $L_3 < 0$) without synchronization between 
SQUID $1$ and $3$ ($\left\langle \eta \right\rangle_\tau > 0.38$).
In this interval, three bifurcations from a quasiperiodic state to another are
visible for the values of $\Omega$ at which $L_1 =L_2 =L_3 =0$.

(vi) $\Omega$ in $(1.216, 1.22]$:
Hyperchaos ($L_1 > L_2 >0$, $L_3 =0$). In this case, the quantity 
$\langle \eta \rangle_\tau$ fluctuates apparently randomly between values which 
either lie below or above the limiting one for intermittent chaos 
synchronization behavior, i.e., $\langle \eta \rangle_\tau =0.38$. By inspection 
of many of the corresponding solutions for the fluxes $\phi_i$ ($i=1,2,3$) in 
this frequency interval, we infer that for $\langle \eta \rangle_\tau < 0.38$ 
the fluxes $\phi_1$ and $\phi_3$ are intermittently synchronized, while for 
$\left\langle \eta\right\rangle_\tau > 0.38$ all the fluxes $\phi_i$ are 
unsynchronized. 

(vii) $\Omega$ in $[1.2, 1.216]$: 
Periodic dynamics ($L_1 =0$, $L_2 < 0$, $L_3 < 0$), with synchronization between 
SQUIDs $1$ and $3$ ($\left\langle \eta\right\rangle_\tau <0.01$).

We illustrate below a typical case of a hyperchaotic state (with accompanied 
intermittent chaos synchronization between SQUIDs $1$ and $3$) and a chaotic
state (with complete chaos synchronization between SQUIDs $1$ and $3$). The
corresponding values of $\Omega$ have been marked by the red horizontal lines A 
($\Omega =1.233$) and B ($\Omega =1.2375$), respectively, in Fig. \ref{fig4}. 
The temporal evolution of the fluxes $\phi_1$ and $\phi_3$ in SQUIDs $1$ and $3$, 
respectively, have been plotted as a function of the normalized temporal variable 
$\tau$ divided by the driving period $T=\Omega/(2\pi)$ in Fig. \ref{fig6}(a) for
$\Omega =1.233$ and Fig. \ref{fig6}(b) for $\Omega = 1.2375$. Blue represents 
the flux in SQUID $1$, $\phi_1$, and red the flux in SQUID $3$, $\phi_3$. In 
Fig. \ref{fig6}(a), the temporal evolution of $\phi_1$ and $\phi_3$ is not 
synchronized although there are some windows in time where synchronization 
occurs. This is typical behavior of hyperchaos with chaos intermittent 
synchronization that will be discussed in more details in the next section. This 
behavior can be confirmed in both ($\phi_1, \phi_2$) and ($\phi_1, \phi_3$) 
plane projections of the trajectory as shown in Fig. \ref{fig6}(c) and (e). 
Complete chaos synchronized can be observed in Fig. \ref{fig6}(b) as the temporal 
evolution of $\phi_1$ and $\phi_3$ practically overlap. The projection of the 
flow onto the ($\phi_1, \phi_3$) plane shown in Fig. \ref{fig6}(f) confirm 
this behavior. The projection of the flow onto the ($\phi_1, \phi_2$) plane 
shown in Fig. \ref{fig6}(d) merely verifies that no synchronization occurs 
between SQUIDs $1$ and $2$.

\section{The parameter space}
As presented in the previous section, based on the time-averaged Euclidean 
distance $\langle \eta \rangle_\tau$, we can observe three main behaviors: 
Complete synchronization, intermittent synchronization and unsynchronized 
solutions. Moreover, by calculating the Lyapunov exponents we identify periodic 
solutions, quasiperiodicity, chaos and hyperchaos. In Fig. \ref{fig7}, a map of 
different dynamical regions based on the combined measurement of  
$\langle \eta \rangle_\tau$ and the maximum Lyapunov exponent are shown in 
($\Omega, \lambda$) (Fig.\ref{fig7} (a)) and ($\Omega, \phi_{ac}$) 
(Fig.\ref{fig7} (b)) parameter space. We observe seven different areas. Periodic 
synchronization (PS) where $L_1 =0, L_2, L_3 <0$ and 
$\langle \eta \rangle _{\tau} < 0.01$, quasiperiodic synchronized 
solutions (QPS) where $L_1 =L_2 =0, L_3 < 0$ and 
$\langle \eta \rangle _{\tau} < 0.01$, periodic unsynchronized solutions 
(PUn)  where $L_1=0, L_2, L_3 <0$ and $0.38 < \langle \eta \rangle _\tau$, 
quasiperiodic unsynchronized solutions (QPUn) where $L_1 =L_2 =0, L_3 <0$ and 
$\left\langle \eta\right\rangle _\tau > 0.38$, chaos synchronization (CS) 
where $L_1 > 0, L_2 =L_3 =0$ and $\left\langle \eta\right\rangle _\tau < 0.01$, 
chaos intermittent synchronization (CI)  where $L_1 > 0, L_2 =L_3 =0$ and 
$0.01 < \left\langle \eta\right\rangle _\tau < 0.38$, and finally hyperchaos 
intermittent synchronization (HCI) where $L_1 > 0, L_2 > 0, L_3 =0$ and 
$0.01 < \left\langle \eta\right\rangle _\tau < 0.38$.

In both ($\Omega, \lambda$) and ($\Omega, \phi_{ac}$) parameter spaces, periodic 
synchronized (PS) and quasiperiodic synchronized (QPS) solutions occupy most of 
the plane. Next, the main dynamical behavior is concentrated in periodic 
unsynchronized (PUn) and quasiperiodic unsynchronized (QPUn) solutions. At the 
boundaries of these regions, we observe chaos synchronization (CS) and hyperchaos 
intermittent synchronization (HCI). It is remarkable that hyperchaos always 
appears with intermittent synchronization of $\phi_1(\tau)$ and $\phi_3(\tau)$. 
We have never observed synchronization or unsynchronized solutions between the 
trimer edges, in the presence of hyperchaos. The opposite is not true. Indeed, 
intermittent synchronization can also be observed in the chaotic regime 
(Fig. \ref{fig7}, blue color (CI)).

Four specific behaviors of non chaotic dynamics are shown in Fig. \ref{fig8} 
for $\phi_{ac} = 0.02$. The time series of the magnetic fluxes $\phi_1(\tau)$ 
and $\phi_3(\tau)$ (Fig. \ref{fig8}(a), left column), for $\Omega = 1.22$ and 
$\lambda=0.16$, show a periodic synchronized solution between SQUID $1$ (red 
color) and SQUID $3$ (blue color). The corresponding time series of $\eta$ 
(Fig. \ref{fig8}(a), right column) which is close to zero also indicates a 
synchronization behavior. When $\Omega = 1.24$ and $\lambda=0.07$, 
$\phi_1(\tau)$ and $\phi_3(\tau)$ oscillate out of phase, with different 
amplitudes, and $\eta$ also oscillates in time with an average value greater 
than $0.38$, an indication for a periodic unsynchronized solution (Fig. 
\ref{fig8}(b), left and right column). For ($\Omega=1.3, \lambda=0.14$) and 
($\Omega=1.255$, $\lambda=0.1018$) the temporal evolution of the system is 
quasiperiodic. The magnetic fluxes of SQUIDs $1$ and $3$ oscillate in time with 
equal and different amplitudes in the left column of Figs.~\ref{fig8}(c) and (d), 
respectively. In the case of synchronization the $\langle \eta \rangle _\tau$ 
quantifier between the trimer edges, i.e., between SQUIDs $1$ and $3$, is less 
than $0.01$ (Figs.~\ref{fig8}(c), right column), while in the unsynchronized 
case $\eta$ has a periodic evolution with multiple frequencies due to the 
quasiperiodic behavior, and $0.38 < \left\langle \eta\right\rangle _\tau$ 
(Fig.~\ref{fig8}(d), right column).

Fig. \ref{fig9} illustrates three other dynamical examples, this time in the 
chaotic regime. For $\Omega = 1.235$ and $\lambda=0.1$ the temporal evolution 
of the system is chaotic and the output magnetic fluxes in SQUIDs $1$ and $3$ 
are identical, as shown in Fig. \ref{fig9} (a), left column. This chaotic 
synchronization is confirmed by the evolution of $\eta$ close to zero in the 
right column. When $\Omega =1.23$ and $\lambda =0.125$ the time series of the 
trimer edges are chaotic but not identical. The three largest Lyapunov exponents 
associated with this evolution are ($0.007,0,0$). Nevertheless, there are some 
windows in time where synchronization occurs (Fig. \ref{fig9}(b), left column). 
This is a chaos intermittent synchronization where $\eta$ evolves, at some 
temporal intervals close to zero (synchronous behavior) while at others with 
large fluctuations between zero and one (Fig. \ref{fig9}(b), right column). The 
same dynamical behavior as in Fig. \ref{fig9}(b) is demonstrated in 
Fig. \ref{fig9}(c) for both $\phi_1(\tau)$ and $\phi_3(\tau)$ (left column) and 
$\eta$ quantifier (right column) where $\Omega =1.23$ and $\lambda=0.11$. 
However, in this case, the associated three largest Lyapunov exponents are 
($0.015,0.002,0$). The system is now hyperchaotic, with more than one positive 
Lyapunov exponent and thus the behavior of the system is characterized as 
hyperchaotic intermittent synchronization.       

\begin{figure}
   \includegraphics[scale=0.5]{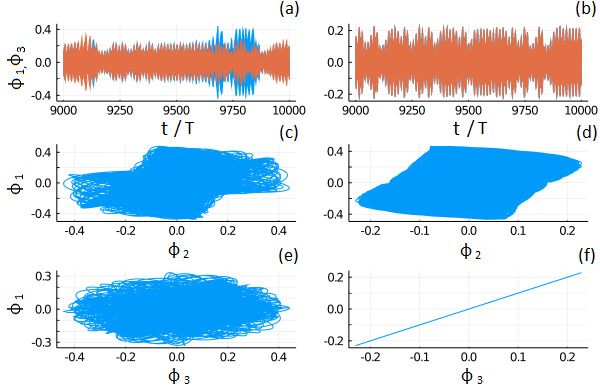}
\caption{Chaos intermittent synchronization for the parameters of Fig. \ref{fig4} 
         and $ \Omega = 1.233$ (red line A). 
         (a) Time series for $\phi_1$ and $\phi_3$. 
         (c) Projection of the flow onto the $\phi_1 - \phi_2$ plane. 
         (e) Projection of the flow onto the $\phi_1 - \phi_3$ plane. 
         (b), (d), and (f) are the corresponding figures in the case of complete
             chaos synchronization for the parameters of Fig. \ref{fig4} and 
             $\Omega =1.2375$ (red line B).}
\label{fig6}
\end{figure}

\begin{figure*}[!t]
   \includegraphics[scale=0.35]{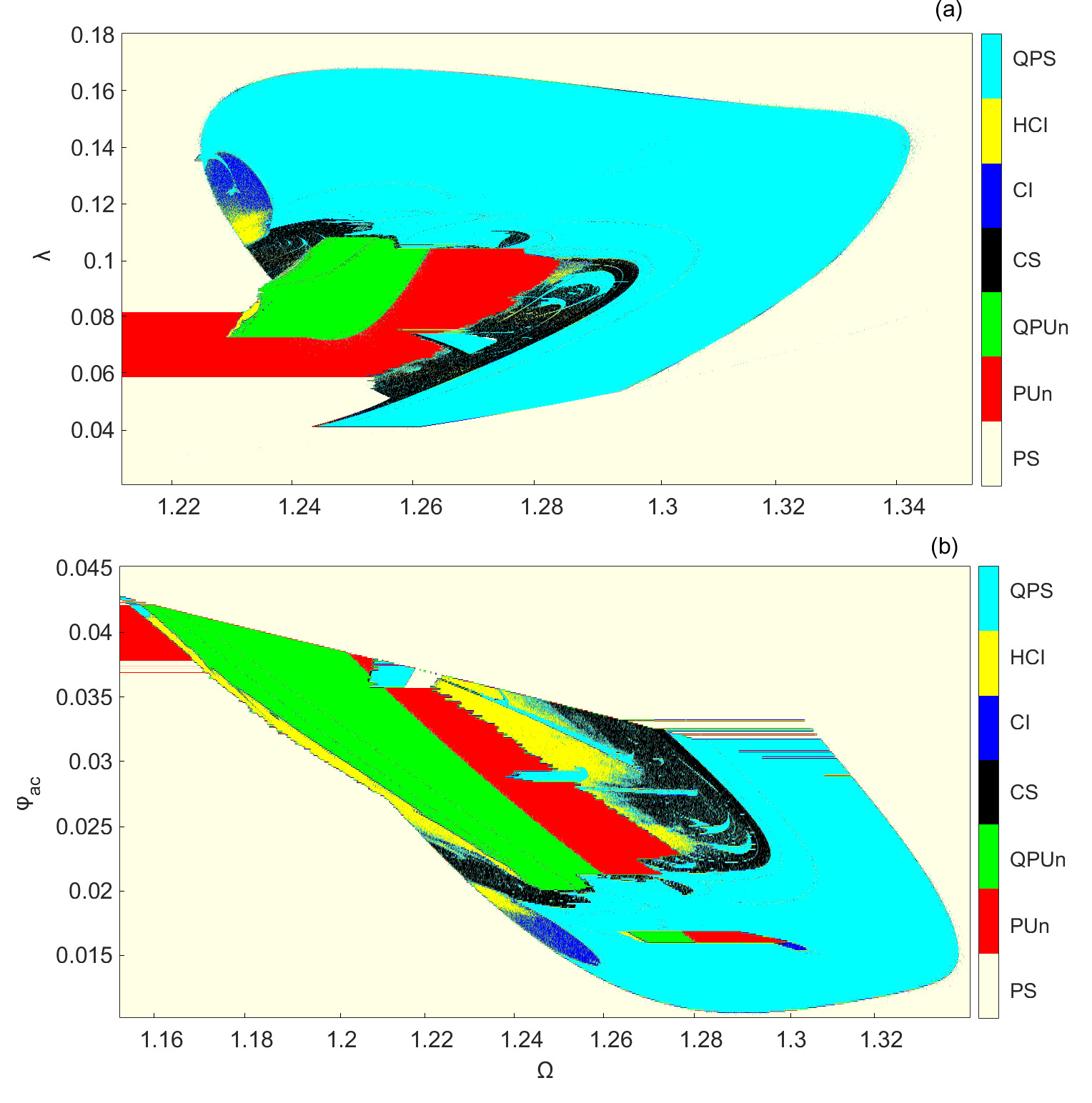}
\caption{Map of different dynamical regions in the 
         (a) ($\lambda, \Omega$) parameter space for $\phi_{ac} = 0.02$, and 
         (b) ($\Omega, \phi_{ac}$) parameter space for $\lambda =0.1075$. 
             Depending on the three largest Lyapunov exponents 
            ($L_1 > L_2 > L_3$) and $\left\langle \eta\right\rangle _\tau$ 
             measurement, we observe seven different areas: 
	     Periodic synchronized solution (PS) where $L_{1}=0, L_2, L_3<0$ and 
             $\left\langle \eta\right\rangle _{\tau} < 0.01$, 
             Quasiperiodic synchronized (QPS) solutions where $L_1=L_2=0, L_3 <0$ 
             and $\left\langle \eta\right\rangle _{\tau}<0.01$,
             Periodic unsynchronized solution (PUn) where $L_{1}=0, L_2, L_3<0$ and 
             $0.38<\left\langle \eta\right\rangle _{\tau}$, 
	     Quasiperiodic unsynchronized solution (QPUn) where 
             $L_{1}=L_{2}=0, L_3<0$ 
             and $\left\langle \eta\right\rangle _{\tau} > 0.38$, 
	     Chaos synchronization (CS) where $L_{1}>0,L_{2}=L_3=0$ and 
             $\left\langle \eta\right\rangle _{t} < 0.01$, 
	     Chaos intermittent synchronization (CI) where $L_{1}>0,L_{2}=L_3=0$ 
             and $0.01<\left\langle \eta\right\rangle _{\tau} < 0.38$ and finally 
	     Hyperchaos intermittent synchronization (HCI)  where 
             $L_{1}>0,L_{2}>0,L_{3}=0$ and 
             $0.01<\left\langle \eta\right\rangle _{\tau} < 0.38$.} 
\label{fig7}
\end{figure*}

\begin{figure}
   \includegraphics[scale=0.5]{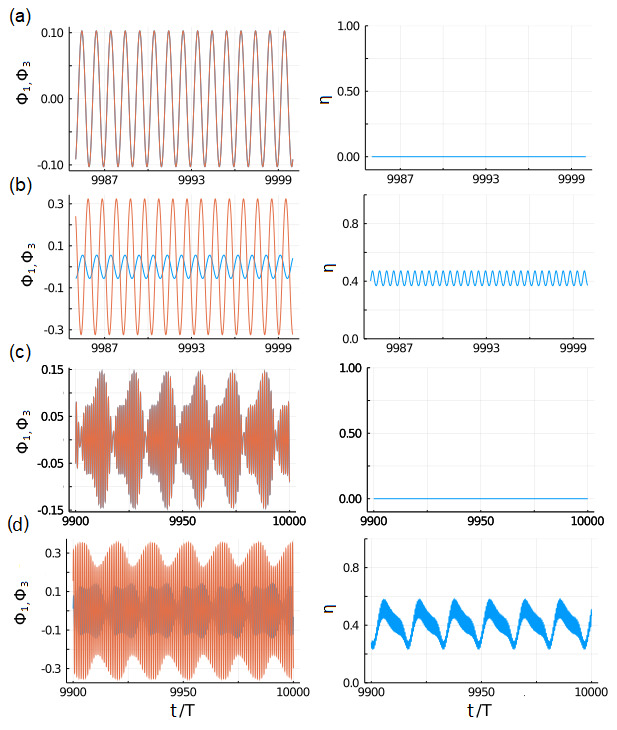}
\caption{
 (a) Periodic synchronized solution (PS) for $\Omega = 1.22$ and $\lambda=0.16$. 
 (b) Periodic unsynchronized (PUn) solution for $\Omega = 1.24$ and 
     $\lambda=0.07$. 
 (c) Quasiperiodic synchronized (QPS) solution for $\Omega=1.3$ and 
     $\lambda=0.14$. 
 (d) Quasiperiodic unsynchronized (QPUn) solution for $\Omega = 1.255$ and 
       $\lambda=0.1018$. 
 In the left column, the fluxes $\phi_1$ and $\phi_3$ through the loops of SQUID
 $1$ (red line) and $3$ (blue line), respectively, are plotted as a function of
 $\tau/T$, where $T=\Omega/(2\pi)$.
 In the right column, the corresponding $\eta$ is plotted as a function of
 $\tau/T$. 
 Other parameters: $\phi_{ac}=0.02$, $\gamma=0.024$, $\phi_{dc}=0$, and 
 $\beta=0.1369$.
} 
\label{fig8}
\end{figure}

\begin{figure}
   \includegraphics[scale=0.5]{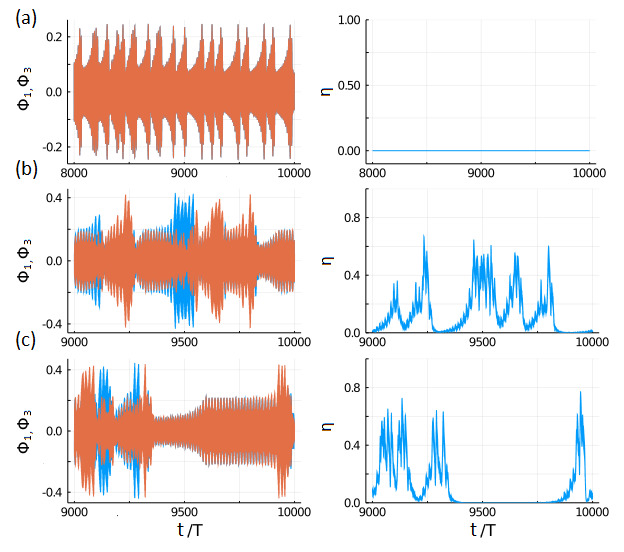}
\caption{
 The fluxes $\phi_1$ and $\phi_3$ through the loops of SQUID $1$ (red line) and 
 $3$ (blue line), respectively, are plotted as a function of $\tau/T$ in the 
 left column, while the corresponding $\eta$s are plotted in the right column.
 (a) Chaos synchronization (CS) for $\Omega = 1.235$ and $\lambda=0.1$. 
 (b) Chaos intermittent synchronization (CI) for $\Omega = 1.23$ and 
     $\lambda=0.125$. 
 (c) Hyperchaos intermittent synchronization (HCI) for $\Omega = 1.23$ and 
     $\lambda=0.11$. 
 Other parameters: $\phi_{ac}=0.02$, $\gamma=0.024$, $\phi_{dc}=0$, and 
 $\beta=0.1369$.
} 
\label{fig9}
\end{figure}

\section{Conclusions}
To summarize, a trimer comprising three identical, magnetically coupled SQUIDs 
is investigated numerically with respect to chaotic synchronization phenomena. 
The SQUID trimer is a superconducting oligomer which serves as a highly complex 
system exhibiting a plethora of nonlinear dynamical effects. In this work, we 
focus on the synchronization between the two SQUIDs at the edges of the trimer 
informed by the corresponding dynamics of the system as a whole. By using 
suitable synchronization measures, like the correlation function and the 
Eulidean distance, we identify the types of synchronization in the relevant 
parameter spaces. Apart from complete chaotic synchronization between SQUIDs $1$
and $3$, we find that the SQUID trimer displays also intermittent chaotic 
synchronization between SQUIDs $1$ and $3$ where intervals of synchronization 
are interrupted by desynchronized activity. Calculations of the full Lyapunov 
exponent spectrum of the system reveal that the way from complete to intermittent 
synchronization is associated to chaos-hyperchaos transitions. In the 
intermittent synchonization case, we observe that the occurrence and the size of 
the intervals of the synchronized/desynchronized chaotic dynamics appear to be 
chaotic themselves. This requires further investigation and will be the subject 
of a future study.
\label{real}

\section{ACKNOWLEDGEMENTS}
This work was supported by the Ministry of Education and Science of the Russian 
Federation in the framework of the Increase Competitiveness Program of NUST 
``MISiS'' (Grant number K4-2018-049).
JH and NL acknowledge support by the General Secretariat for Research and 
Technology (GSRT) and the Hellenic Foundation for Research and Innovation 
(HFRI) (Code No. 203).
\section{DATA AVAILABILITY}
The data that support the findings of this study are available
from the corresponding author upon reasonable request.

\end{document}